\documentstyle[12pt,fleqn]{article}
\def\beq{\begin{equation}}
\def\eeq{\end{equation}}
\def\0{\otimes}
\def\6{\langle}
\def\9{\rangle}
\def\Ps{{\mit\Psi}}

\begin{document}

\begin{center}
{\Large {\bf Karl Popper and the Copenhagen Interpretation}}\bigskip

Asher Peres\\[2mm]
{\sl Department of Physics, Technion---Israel Institute of Technology,
32\,000 Haifa, Israel}\\[10mm]

{\bf Abstract}\end{center}

\begin{quote} Popper conceived an experiment whose analysis led to a
result that he deemed absurd. Popper wrote that his reasoning was based
on the Copenhagen interpretation and therefore invalidated the latter.
Actually, Popper's argument involves counter\-factual reasoning and
violates Bohr's complementarity principle. The absurdity of Popper's
result only confirms Bohr's approach.\\[12mm]
\end{quote}

\hspace*{\fill}\parbox{62mm}{I called thee to curse mine enemies, and,
behold, thou hast altogether blessed them.}\smallskip

\hspace*{\fill}{\it Numbers\/} 24:10\bigskip

The emergence of quantum mechanics led to considerable progress in our
understanding of physical phenomena. However, it also led to serious
misconceptions. In my current work as a theoretical physicist, I
recently examined a conceptual experiment that was proposed some time
ago by Karl Popper (1982). Its feasibility was challenged by Collett and
Loudon (1987) who claimed that such an experiment would be inconclusive.
Nevertheless, an actual experiment is currently under way (Kim and Shih,
1999). The rigorous theoretical analysis of these experiments is quite
intricate and I shall only briefly outline it here. Most of the present
article is an attempt to analyze the meaning of what Popper wrote and to
understand his way of reasoning. I found it most surprising when I read
the original argument in his book.

Popper's experiment is a variant of the one considered long ago by
Einstein, Podolsky, and Rosen (1935): a source S emits pairs of
particles having a broad angular distribution but precisely opposite
momenta,

\beq {\bf p}_1+{\bf p_2}=0. \eeq 
The example given by Popper is that of pairs of photons emitted by the
decay of positronium at rest. Actually, the wavelength of gamma rays
emitted by positronium is much too short for realizing Popper's
experiment, but pairs of photons resulting from parametric
down-conversion in a nonlinear crystal (Kim and Shih, 1999) are suitable
for that purpose: these photons have precisely correlated (though not
opposite) momenta, and this is all we need. If we wish, we can refer our
calculations to a Lorentz frame moving with a constant velocity
$c\,({\bf p}_1+{\bf p_2})/(E_1+E_2)$, so that Eq.~(1) holds in that
frame.

Note that Eq.~(1) seems to conflict with the quantum ``uncertainty
principle.'' Popper writes ``we consider pairs of particles that move
in opposite directions along the positive and negative $x$-axis.'' If
these were classical particles, opposite momenta would indeed lead to
opposite positions. However, the quantum dynamical variables in Eq.~(1)
do not commute with $({\bf q_1}+{\bf q_2})$. For the components along
any axis, we have uncertainty relations

\beq \Delta(p_1+p_2)\;\Delta(q_1+q_2)\ge\hbar, \eeq
which set a limit on how precisely opposite the positions of the
particles will be observed. This issue was analyzed by Collett and
Loudon (1987) who came to the conclusion that Popper's experiment
(described below) could not give conclusive results. This is just one
example of how hazardous it is to use classical reasoning when we
discuss quantum phenomena. I shall return to this point later. 

However, it is no less hazardous to make heuristic use of the
``uncertainty principle'' in order to draw quantitative conclusions.
What must be done in case of doubt is to write the Schr\"odinger
equation that describes the physical situation, and to derive rigorously
unambiguous results. As will be shown below, the analysis of Popper's
experiment is much subtler than either Popper, or Collett and Loudon,
were inclined to think.

Popper's proposed experiment proceeds as follows: two observers, whom I
shall call Alice and Bob in accordance with current practice in quantum
information theory, are located on opposite sides of the source, with
arrays of detectors as shown in Figure~1. Alice can place an opaque
screen with a narrow slit of width $a$ in the way of her photons, so
that those passing through the slit are diffracted by an angle of the
order of $\lambda/a$, where $\lambda$ is the wavelength of the photons.
The narrower the slit, the wider is the scattering angle. 

On this point, Popper writes that ``the wider scattering angles go with
a narrower slit, according to the Heisenberg relations.'' Actually, the
diffraction angle $\lambda/a$ is a well known result of classical
optics. The wavelength of the photons, which is the quantity that we can
actually measure, is related to their momentum by the relation
$\lambda=h/p$, which readily follows from Einstein's equation for the
photoelectric effect, $E=h\nu$. The latter predates Heisenberg's
uncertainty principle by more than 20 years. Still before Heisenberg, it
was de~Broglie's bold intuition to extend the relation $\lambda=h/p$ to
massive particles, and in that case $\lambda$ is called the de~Broglie
wavelength. However, the issue is not just one of misappropriation of
credit. Here, Popper wanted to invoke Heisenberg's ``uncertainty''
because he had in mind that the detection of a particle that had passed
through Alice's slit was a measurement of the $y$-coordinate of that
particle at it passed through the slit, and therefore also a virtual
measurement of the position of the other particle, since the two had
precisely opposite directions. Let us examine Popper's text:

\begin{quote} According to the EPR argument, we have measured $q_y$ for
both particles \ldots\ with the precision $\Delta q_y$ [$\equiv a$]
\ldots\ We can now calculate the $y$-coordinate of the [other] particle
with approximately the same precision \dots\ We thus obtain fairly
precise {\it `knowledge'\/} about the $q_y$ position of this
particle---we have `measured' its position indirectly. And since it is,
according to the Copenhagen interpretation, our {\it knowledge\/} which
is described by the theory \ldots\ we should expect that the momentum of
the [second] beam scatters as much as that of the beam that passes
through the slit \ldots

To sum up: if the Copenhagen interpretation is correct, then any
increase of our {\it mere knowledge\/} of the position \ldots\ of the
particles \ldots\ should increase their scatter \ldots\end{quote}

\noindent The italics that appear in the above excerpt are those in the
book. Popper refrains from openly saying that the above prediction is
absurd (as it obviously is). He only says that he is ``inclined to
predict'' that the test will decide against the Copenhagen
interpretation. On this, I have several comments. 

First, is not at all clear why Popper associates this absurd prediction
(particle scatter due to potential knowledge by an observer) with the
Copenhagen interpretation. This is another example of credit
misappropriation, much worse than having quoted Heisenberg instead of
Einstein or de~Broglie. Whatever the ``Copenhagen interpretation'' is (a
point that I shall discuss later), it is reasonable to expect that it is
somehow related to the views expressed by Niels Bohr. However, Popper
himself wrote explicitly that his proposed experiment was an extension
of the argument of Einstein, Podolsky, and Rosen (1935). It is well
known that their argument was promptly criticized by Bohr (1935). I find
it quite remarkable that an opinion which is diametrically opposite to
Bohr's be called the ``Copenhagen interpretation.''

I also have other, more serious objections to the terminology used in
the passage quoted above. In particular, I take exception to the phrase
``we have measured $q_y$'' of some particle. Here however, my criticism
is not aimed at Popper because we are all guilty of occasionally talking
like that. This is a misleading language, as explained long ago by
Kemble (1937):

\begin{quote} We have no satisfactory reason for ascribing objective
existence to physical quantities as distinguished from the numbers
obtained when we make the measurements which we correlate with them.
There is no real reason for supposing that a particle has at every
moment a definite, but unknown, position which may be revealed by a
measurement of the right kind, or a definite momentum which can be
revealed by a different measurement. On the contrary, we get into a maze
of contradictions as soon as we inject into quantum mechanics such
concepts carried over from the language and philosophy of our
ancestors\ldots\ It would be more exact if we spoke of ``making
measurements'' of this, that, or the other type instead of saying that
we measure this, that, or the other ``physical quantity.'' \end{quote}

\noindent Terms that Popper used, such as ``knowledge of the
$y$-coordinate \ldots\ or the $q_y$ position of this particle'' are
flagrant (and admittedly quite common) abuses of an improper language.
When we are discussing quantum theory, we should refrain from using
classical terminology---or at least be aware that we do so at our own
risk. 

In classical mechanics, a particle has (ideally) a precise position and
a precise momentum. We can in principle measure them with arbitrary
accuracy and thereby determine their numerical values. In quantum
mechanics, a particle also has a precise position and a precise
momentum. However, the latter are mathematically represented by
self-adjoint operators in a Hilbert space, not by ordinary numbers.
Their nature is quite different from that of the classical position and
momentum. In the early quantum literature, operators were called
$q$-numbers, while plain numbers were $c$-numbers (Dirac, 1926).
Likewise, to avoid confusion, we should have used in quantum theory
names such as $q$-position and $q$-momentum, while the corresponding
classical dynamical variables would have been called $c$-position and
$c$-momentum. If such a distinction had been made, it would have helped
to prevent much of the present confusion about quantum theory. It is the
imperfect translation from the $q$-language to the $c$-language that led
to the unfortunate introduction of the term ``uncertainty'' in that
context.

We may note, incidentally, that the theory of relativity did not cause
as much misunderstanding and controversy as quantum theory, because
people were careful to avoid using the same nomenclature as in
nonrelativistic physics. For example, elementary textbooks on relativity
theory distinguish ``rest mass'' from ``relativistic mass'' (hard core
relativists call them simply ``mass'' and ``energy'').

The criticism above was aimed at the terminology used by Popper in
proposing his experiment. Now, it is time to analyze the substance.
First, we have to find out how precisely the two particles of each pair
will be aligned opposite to each other, in spite of the uncertainty {\it
relation\/} in Eq.~(2). Note that, contrary to the so-called
``uncertainty principle'' which is an ill defined concept and has only a
heuristic meaning, Eq.~(2) is a rigorous mathematical consequence of the
quantum formalism. It puts a lower bound on the product of the standard
deviations of the results of a large number of measurements performed on
identically prepared systems. Each one of these measurements is assumed
to have perfect accuracy (any experimental inaccuracy would have to be
added to the quantum dispersion). There is no ``uncertainty''
connotation here, unless this uncertainty merely refers to future
outcomes of potential, perfectly accurate measurements that may be
performed on such systems (Ballentine, 1970).

A long calculation (to be published separately) is needed to estimate
how precise is the angular alignment of two particles emitted with
opposite momenta. Actually, what Eq.~(2) says is that if an ensemble of
pairs of particles is prepared in such a way that $(p_1+p_2)$ is sharp,
then the positions of the points halfway between the particles are very
broadly distributed. It says nothing on the angular alignment of distant
particles. On that issue, a detailed calculation shows that if one
particle is found in the direction given by polar and azimuthal angles
$\theta$ and $\phi$, then the other will be found very nearly in the
opposite direction, with angles $\pi-\theta$ and $\phi\pm\pi$,
respectively. The allowed deviation from perfect alignment is too small
to be of any consequence in the present discussion.

It is therefore correct to assume, as Popper did, that if a particle is
detected behind Alice's slit, and if an identical slit were placed by
Bob in a symmetric position, then Bob would definitely detect the other
particle of that pair there. However, this does not mean that Bob's
knowledge creates a ``virtual slit'' through which his particles are
diffracted by the same angle $\lambda/a$. Bob's knowledge has no
physical consequence because it is manifestly counter\-factual. This
can easily be seen by considering other counter\-factual experiments.
For example, Bob also knows, after he was informed by Alice of what she
found, that if he had placed a slit of width $a/2$ at a position whose
distance from the source is one half of the distance of Alice's slit,
then he would have detected his particle within that slit with
certainty. In that case, his ``virtual slit'' is narrower, and
therefore the diffraction angle is wider by a factor~2. In brief, we can
imagine infinitely many such counter\-factual experiments (which are
mutually exclusive, of course), and each one of these conceptual slits
leads to a different observable diffraction angle, which is absurd. 

There is no doubt that Popper was right when he was ``inclined to
predict'' that the test would give a negative result. However, Popper
concluded that ``the test decides against the Copenhagen
interpretation'' and this assertion requires further scrutiny. What is,
indeed, the Copenhagen interpretation? There seems to be at least as
many different Copenhagen interpretations as people who use that term,
probably there are more. For example, in two classic articles on the
foundations of quantum mechanics, Ballentine (1970) and Stapp (1972)
give diametrically opposite definitions of ``Copenhagen.'' There is no
real conflict between Ballentine and Stapp on how to understand quantum
mechanics, except that one of them calls Copenhagen interpretation what
the other considers as the exact opposite of the Copenhagen
interpretation. I shall now explain my own Copenhagen interpretation. It
relies on articles written by Niels Bohr. Whether or not you agree with
Bohr, he is the definitive authority for deciding what is genuine
Copenhagen.

Quantum mechanics provides statistical predictions for the results of
measurements performed on physical systems that have been prepared in
specified ways (Peres, 1995). (I hope that everyone agrees at least with
that statement. The only question here is whether there is more than
that to say about quantum mechanics.) The preparation of quantum systems
and their measurement are performed by using laboratory hardware which
is described in {\it classical\/} terms. If you have doubts about that,
just have a look at any paper on experimental physics. The necessity of
using a classical terminology was emphasized by Bohr (1949) whose
insistence on this point was very strict: 

\begin{quote} However far the [quantum] phenomena transcend the scope of
classical physical explanation, the account of all evidence must be
expressed in classical terms. The argument is simply that by the word
`experiment' we refer to a situation where we can tell others what we
have done and what we have learned and that, therefore, the account of
the experimental arrangement and the results of the observations must be
expressed in unambiguous language with suitable application of the
terminology of classical physics.\end{quote}

\noindent The keywords in that excerpt are: {\it classical terms \ldots\
unambiguous language \ldots\ terminology of classical physics\/}. Bohr
did {\it not\/} say that there are in nature classical systems and
quantum systems. There are physical systems for which we may use a
classical description or a quantum description, according to
circumstances, and with various degrees of approximation. It is
according to our assessment of the physical circumstances that we decide
whether the $q$-language or the $c$-language is appropriate. Physics is
not an exact science, it is a science of approximations. Unfortunately,
Bohr was misunderstood by some (perhaps most) physicists who were unable
to make the distinction between language and substance, and he was also
misunderstood by philosophers who disliked his positivism.

It is remarkable that Bohr never considered the measuring process as a
dynamical interaction between an apparatus and the system under
observation. Measurement had to be understood as a primitive notion.
Bohr thereby eluded questions which caused considerable controversy
among other authors (Wheeler and Zurek, 1983). Bohr willingly admitted
that any intermediate systems used in the measuring process could be
treated quantum mechanically, but the {\it final\/} instrument always
had a purely classical description (Bohr, 1939):

\begin{quote} In the system to which the quantum mechanical formalism is
applied, it is of course possible to include any intermediate auxiliary
agency employed in the measuring process [but] some ultimate measuring
instruments must always be described entirely on classical lines, and
consequently kept outside the system subject to quantum mechanical
treatment.\end{quote}

Yet, a quantum measurement is not a supernatural process. Measuring
apparatuses are made of the same kind of matter as everything else and
they obey the same physical laws. It therefore seems natural to use
quantum theory in order to investigate their behavior during a
measurement. This was first attempted by von Neumann (1932) in his
treatise on the mathematical foundations of quantum theory. In the last
section of that book, as in an after\-thought, von Neumann represented
the apparatus by a single degree of freedom whose value was correlated
to that of the dynamical variable being measured.  Such an apparatus is
not, in general, left in a definite pure state, and does not admit a
classical description. Therefore, von Neumann introduced a second
apparatus which observes the first one, and possibly a third apparatus,
and so on, until there is a final measurement, which is {\it not\/}
described by quantum dynamics and has a definite result (for which
quantum mechanics can only give statistical predictions). The essential
point that was suggested, but not proved by von Neumann, is that the
introduction of this sequence of apparatuses is irrelevant: the final
result is the same, irrespective of the location of the ``cut'' between
classical and quantum physics. (At this point, von Neumann also
speculated that a final step would involve the consciousness of the
observer---a rather bizarre statement in a mathematically rigorous
monograph.)

These different approaches of Bohr and von Neumann were reconciled by
Hay and Peres (1998), who introduced a dual description for the
measuring apparatus. It obeys quantum mechanics while it interacts with
the  system under observation, and then it is ``dequantized'' and is
described by a classical Liouville density, which provides the
probability distribution for the results of the measurement.
Alternatively, the apparatus may always be treated by quantum mechanics,
and be measured by a second apparatus which has such a dual description.
Hay and Peres showed that these two different methods of calculation
give the same result, provided that the measuring apparatus satisfies
appropriate conditions (otherwise, it is not a valid measuring
apparatus).

The other fundamental feature of Bohr's presentation of quantum theory
is the principle of complementarity, which asserts that when some types
of predictions are possible, others are not, because they are related to
mutually incompatible experiments. For example, in the situation
described by Einstein, Podolsky, and Rosen (1935), the choice of the
experiment performed on the first system determines the type of
prediction that can be made for the results of experiments performed on
the second system (Bohr, 1935). 

In Popper's experiment, Bob can predict what would have happened if he
had placed slits of various sizes at various positions, or no slit at
all. However, all these possible setups are mutually incompatible. In
particular, if Bob puts no slit at all, the result he obtains is not the
one he would have obtained if he had put a slit. Counter\-factual
experiments need not have consistent results (Peres, 1978).

Note that Bohr did not contest the validity of counterfactual reasoning.
He wrote (Bohr, 1935): 

\begin{quote}Our freedom of handling the measuring instruments is
characteristic of the very idea of experiment \ldots\ we have a
completely free choice whether we want to determine the one or the other
of these quantities \ldots\end{quote}

\noindent Thus, Bohr found it perfectly legitimate to consider
counter\-factual alternatives: observers have free will and can
arbitrarily choose their experiments. However, each experimental setup
must be considered separately. In particular, no valid conclusion can be
drawn from the comparison of possible results of mutually incompatible
experiments. Bohr was sometimes accused of being elusive, because his
approach does not provide answers to questions in which people may be
interested. There are indeed questions that seem reasonable but do not
correspond to any conceivable experiment: quantum theory has no
obligation to answer meaningless questions.

To conclude this article, let me report the result of a rigorous
analysis of Popper's experimental setup, where only Schr\"odinger's
equation is used, without invoking any controversial interpretation. The
irony of the answer is that Bob does observe a diffraction broadening,
as if he had a virtual slit! However, that slit is not located between
him and the source, but is precisely located where Alice's real slit is,
and is indeed identical to it. An experiment similar to Popper's
proposal was actually performed by Strekalov {\it et~al.\/} (1995), who
used a double slit, so that Bob had a virtual double slit, producing a
neat interference pattern, not only a diffraction broadening. Figure~2
is a simplified sketch of that experiment. Its complete theoretical
analysis involves advanced concepts of quantum optics and is quite
intricate. I shall now give a brief outline of the theory, based on
Schr\"odinger's equation.

The only ``knowledge'' needed in the analysis of the experiment is the
factual one, on the preparation and observation procedures. That
knowledge is formally encapsulated in the Hilbert-space vectors
$|\Ps_0\9$ and $|\Ps_d\9$, whose coordinate-space representation is
localized in the source of particles and in the detectors that were
excited, respectively. (These vectors are also known as ``quantum
states.'') Schr\"odinger's equation asserts that the initial vector
$|\Ps_0\9$ evolves in time, as long as there is no detection event,
according to a unitary transformation  

\beq |\Ps_0\9\to|\Ps_t\9=U_t\,|\Ps_0\9,\eeq
where $U_t=e^{-iHt/\hbar}$ for a time-independent Hamiltonian $H$. In
the present case, the double slit can be represented by an infinite
potential in $H$, or by an equivalent boundary condition.

Born's rule (which makes the connection between the quantum formalism
and observed probabilities of macroscopic events) asserts that the
probability that a particular pair of detectors will ``click'' at time
$t$ is $P=|\6\Ps_d,\Ps_t\9|^2$, where the symbol $\6u,v\9$ denotes the
scalar product of two vectors, $|u\9$ and $|v\9$. We thus have (Peres,
1995)

\beq P=|\6\Ps_d,U_t\,\Ps_0\9|^2  
  \equiv|\6U_t^\dagger\,\Ps_d,\Ps_0\9|^2,\eeq
where $U_t^\dagger=U_{-t}$ is the unitary operator for the
time-reversed dynamics. It may be practically impossible to realize
experimentally that reversed dynamics, but it is legitimate to perform
the calculation of the ordinary dynamics by proceeding backwards,
starting at the detectors and ending at the source. In the present case,
this is indeed much easier, because $|\Ps_0\9$ is entangled and has to
satisfy Eq.~(1), while

\beq |\Ps_d\9=|\psi_1\9\0|\psi_2\9, \eeq
is a tensor product of two vectors, whose coordinate-space
representations are well separated, since they are localized in the two
detectors. Moreover, the Hamiltonian is the sum of those of the two
particles, since the latter do not interact after they leave the source.
Therefore the unitary evolution also factorizes: $U_{-t}=U_1\0U_2$. We
thus propagate $|\psi_1\9$ and $|\psi_2\9$ from the detectors toward the
source. We have to compute

\beq P=|\6\Ps_0,(U_1\psi_1\0U_2\psi_2)\9|^2. \eeq
Now, since $|\Ps_0\9$ satisfies Eq.~(1), the only contribution to $P$
comes from components of $|U_1\psi_1\0U_2\psi_2\9$ with opposite momenta
that also satisfy Eq.~(1). This is illustrated in Figure~2. For example,
if we record all the detections on Bob's side that are in coincidence
with one particular detector of Alice, then Bob will observe an ordinary
double-slit interference pattern, generated by a ``virtual''
double-slit, that actually is Alice's real slit.

Note that it is necessary, for such an observation to be possible, that
the region of the nonlinear crystal from where the rays emerge be very
broad (Hong and Mandel, 1985) and the emergence point be undetermined.
Likewise, if the experiment were done with positronium as Popper
originally suggested, the positronium ought to be prepared with $\Delta
y$ much larger than the distance between the slits. Expressed in an
informal language, the requirement is that each one of the two photons
that pass through both slits must also originate in {\it both\/} regions
of the source. This demand is similar to the conditions required for the
Pfleegor and Mandel (1967) experiment, where a single photon originates
from two different lasers and gives rise to first order interference. A
similar analysis also applies to Popper's original experiment with a
single slit (however, it would be more difficult to draw for it a figure
like Figure~2).

In summary, according to the Copenhagen interpretation, as Bohr
apparently understood it, quantum theory is not a description of
physical reality. It also does not deal with anthropomorphic notions
such as knowledge or consciousness. All it does is to provide correct
answers to meaningful questions about experiments done with physical
systems.

\begin{center}{\bf Acknowledgments}\end{center}

I am grateful to Rainer Plaga for bringing Popper's experiment to my
attention, and to Amiram Ron for clarifying discussions. This work was
supported by the Gerard Swope Fund and the Fund for Encouragement of
Research.\clearpage

\begin{center}{\bf References}\end{center}\frenchspacing

\begin{description}

\item Ballentine, L. E. (1970) `The Statistical Interpretation of
Quantum Mechanics' {\it Reviews of Modern Physics\/} {\bf 42}, 358--381.

\item Bohr, N. (1935) `Can Quantum-Mechanical Description of Physical
Reality be Considered Complete?' {\it Physical Review\/} {\bf 48},
696--702.

\item Bohr, N. (1939) `The causality problem in modern physics' in {\it
New Theories in Physics\/} (Paris: International Institute of
Intellectual Cooperation) pp.~11--45.

\item Bohr, N. (1949) `Discussion with Einstein on Epistemological
Problems in Atomic Physics' in P.~A.~Schilpp (ed.) {\it Albert Einstein,
Philosopher-Scientist\/} (Evanston: Library of Living Philosophers)
pp.~201--241. 

\item Collett, M. J., and Loudon, R. (1987) `Analysis of a proposed
crucial test of quantum mechanics' {\it Nature\/} {\bf326}, 671--672.

\item Dirac, P. A. M. (1926) `Quantum mechanics and a preliminary
investigation of the hydrogen atom' {\it Proceedings of the Royal
Society A\/} (London) {\bf110}, 561--569.

\item Einstein, A., Podolsky, B., and Rosen, N. (1935) `Can
Quantum-Mechanical Description of Physical Reality be Considered
Complete?' {\it Physical Review\/} {\bf 47}, 777--780.

\item Hay, O., and Peres, A. (1998) `Quantum and classical descriptions
of a measuring apparatus' {\it Physical Review A\/} {\bf 58}, 116--122.

\item Kemble, E. C. (1937) {\it The Fundamental Principles of Quantum
Mechanics\/} (New York: McGraw-Hill, reprinted by Dover) pp.~243--244.

\item Kim, Y. H., and Shih, Y. H. (1999) `Experimental realization of
Popper's experiment: violation of the uncertainty principle?' {\it
Fortschritte der Physik\/} (in press).

\item Hong, C. K., and Mandel, L. (1985) `Theory of parametric frequency
down conversion of light' {\it Physical Review A\/} {\bf 31},
2409--2418.

\item Peres, A. (1978) `Unperformed experiments have no results' {\it
American Journal of Physics\/} {\bf46}, 745--747.

\item Peres, A. (1993) {\it Quantum Theory: Concepts and Methods\/}
(Dordrecht: Kluwer Academic Publishers).

\item Pfleegor, R. L., and Mandel, L. (1967) `Interference of
independent photon beams' {\it Physical Review\/} {\bf159}, 1084--1088.

\item Popper, K. R. (1982) {\it Quantum Theory and the Schism in
Physics\/} (London: Hutchinson) pp.~27--29.

\item Stapp, H. P. (1972) `The Copenhagen Interpretation' {\it American
Journal of Physics\/} {\bf 40}, 1098--1116.

\item Strekalov, D. V., Sergienko, A. V., Klyshko, D. N., and Shih, Y.
H. (1995) `Observation of two-photon ``ghost'' interference and
diffraction' {\it Physical Review Letters\/} {\bf74}, 3600--3603.

\item von Neumann, J. (1932) {\it Mathematische Grundlagen der
Quantenmechanik\/} (Berlin: Sprin\-ger); transl.\ by R.~T.~Beyer (1955)
{\it Mathematical Foundations of Quantum Mechanics\/} (Princeton:
Princeton University Press).

\item Wheeler, J. A., and Zurek, W. H., editors (1983) {\it Quantum
Theory and Measurement\/} (Princeton: Princeton University Press). 
\end{description}\nonfrenchspacing\vfill

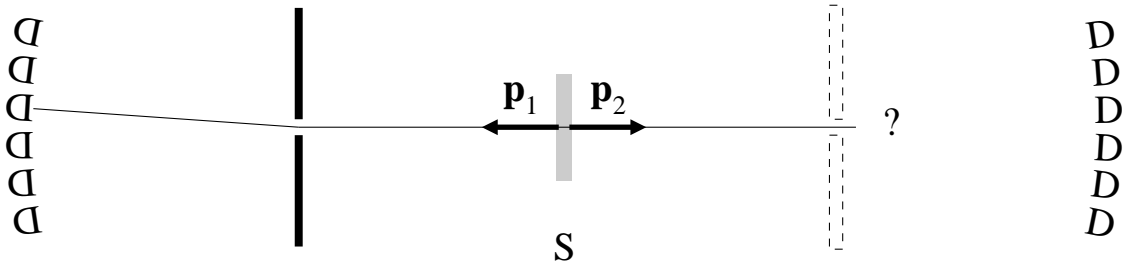
\noindent FIGURE 1. \ Popper's conceptual experiment. A pair of photons
with opposite momenta is emitted by the source S. Alice's detectors are
on the left, those of Bob on the right.\bigskip

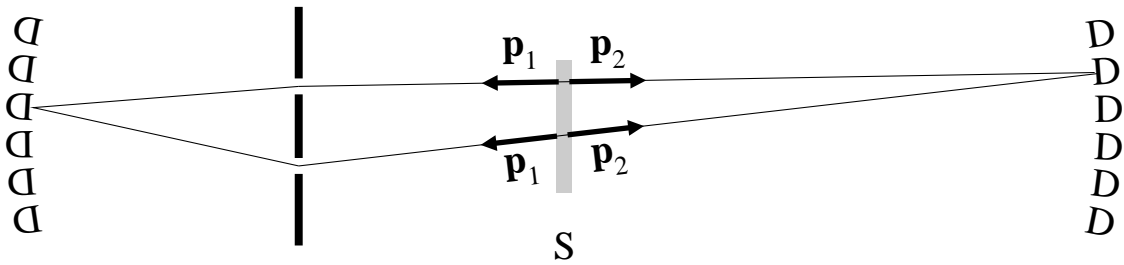
\noindent FIGURE 2. \ Simplified sketch of the experiment of Strekalov
{\it et~al.\/} (1995). The figure shows a {\it single\/} pair of photons
with opposite momenta, emitted by the source S. When many such pairs are
detected in coincidence, interference patterns appear on both sides.

\end{document}